\documentstyle[12pt,epsf]{article}
\def\eqalign#1{\,\vcenter{\openup1\jot \mathsurround=0pt 
        \ialign{\strut \hfil$\displaystyle{##}$&$
        \displaystyle{{}##}$\hfil\crcr#1\crcr}}\,}
\def\etal{\hbox{\it et. al.}}
\def\VEV#1{\left\langle #1\right\rangle}
\def\holdtheequation{\arabic}
\def\sectioneq{\def\holdtheequation{\thesection.\arabic}\let
      \section={\section\setcounter{equation}{0}}\setcounter
      {equation}{0}\def\theequation{\holdtheequation{equation}}}

\def\auto{\eqno(\refstepcounter{equation}\theequation)}
\def\zp{Z.\ Phys.\ }
\def\np{Nucl.\ Phys.\ }
\def\pr{Phys.\ Rev.\ }
\def\prl{Phys.\ Rev.\ Lett.\ }
\def\pl{Phys.\ Lett.\ }
\renewcommand{\thefootnote}{\fnsymbol{footnote}}
\def\mainhead#1{\setcounter{equation}{0}\addtocounter{section}{1}
     \vbox{\begin{center}\large\bf #1\end{center}}\nobreak\par}
\def\tablehead#1{\vbox{\begin{center}\large\bf #1\end{center}}\nobreak\par}
\def\subhead#1{\vbox{\medskip\noindent \bf #1}\nobreak\par}
\def\rf#1#2#3{{\bf #1}, #2 (19#3)}

\renewcommand{\arraystretch}{1.5}
\sectioneq
                                  
\begin{document} \begin{titlepage} \rightline{\vbox{\halign{#\hfil \cr
\normalsize ANL-HEP-PR-98-08 \cr
\normalsize JLAB-THY-98-08 \cr
\normalsize hep-ph/9803351 \cr
\normalsize March 13, 1998 \cr}}}
\vspace{0.5cm} 
\begin{center} 

\Large 
{\bf $x$-Dependent Polarized Parton Distributions}
\medskip\medskip\medskip

\normalsize
Lionel E. Gordon$^{a,b,c}$ Mehrdad Goshtasbpour$^{d,e}$ and 
Gordon P. Ramsey$^{f,c}$
\end{center}

\renewcommand{\arraystretch}{1.0}
\hspace{1.5cm}
{\it a) Thomas Jefferson National Lab, Newport News, VA 23606}

\hspace{1.5cm}
{\it b) Hampton University, Hampton, VA 23668}

\hspace{1.5cm}
{\it c) Argonne National Lab, Argonne, IL 60439}
\footnote{Work supported in part by the U.S. Department of Energy, Division of
High Energy Physics, Contract W-31-109-ENG-38.}
 
\hspace{1.5cm}
{\it d) Shahid Beheshti University, Tehran, Iran}

\hspace{1.5cm}
{\it e) Center For Theoretical Physics and Mathematics, AEOI, Tehran, Iran}

\hspace{1.5cm}
{\it f) Loyola University, Chicago, IL 60626}

\renewcommand{\arraystretch}{1.5}
\begin{abstract}
Using QCD motivated and phenomenological considerations, we construct $x$
dependent polarized parton distributions, which evolve under GLAP evolution,
satisfy DIS data and are within positivity constraints. Each flavor is done
separately and the overall set can be used to predict polarization asymmetries
for various processes. We perform our NLO analysis strictly in $x$ space,
avoiding difficulties in moment inversion. Small-$x$ results and other
physical considerations are discussed.
\end{abstract}


\renewcommand{\thefootnote}{\arabic{footnote}} \end{titlepage}

\mainhead{I. Introduction}

In light of recent polarized deep-inelastic-scattering data, there has been
considerable interest in generating $x$-dependent parton distributions for
the spin-dependent case. Physics results have been extracted from the 
integrated structure functions [1-3]. The results indicate that there is still
considerable uncertainty in the fraction of spin carried by the gluons and sea 
quarks. Each of the analyses rely on certain assumptions to model the 
polarized distributions. One important way to test these assumptions is to
generate the $x$-dependent distributions and predict spin observables, such as
the structure functions, $g_1$, for the proton, neutron and deuteron, hard
scattering cross sections for polarized hadronic collisions and hadronic
production of pions, Kaons and heavy quark flavors. The structure function
measurements of $g_1$ have been made, but the distributions at small-$x$ are
still quite uncertain.
 
There are various sets of $x$-dependent polarized distributions which extract
the unknown parameters from assumptions about the data [3-9].
Most of these are consistent with the $x$-dependent data, but do not
adequately address the physical questions of compatibility with the integrated
data (and hence the spin fractions of the partons) and the positivity
constraints for each flavor. The usual approach is to fit the polarized data
directly with a given parametrization, and then check the integrals for 
agreement with the extracted fractions of spin carried by the quark flavors
(or set normalizations to fit the spin fractions). Often, either the valence
is not considered separately or the flavor dependence of the sea is not
considered. 

Our approach is to establish a reasonable set of flavor-dependent distributions
at an initial $Q_0^2$, motivated by physical constraints and data, then evolve
to arbitrary $Q^2$ for use in predicting polarized observables. This approach
is unique in many aspects. The distributions begin with information from the 
integrated distributions and impose normalization and positivity constraints
(all flavors, valence and sea) to ensure that all of the spin information
extracted from data is explicitly contained in the $x$-dependent results.
We generate polarized parton distributions from the unpolarized distributions
and well defined suitable assumptions, derived from the most recent polarized
deep-inelastic-scattering (PDIS) data sets available [10-15].

For the polarized sea, we assume a broken SU(3) model, to account for mass
effects in polarizing the sea. Our models separate out all flavors
in the valence and sea for a complete analysis of the flavor dependence of
the spin fractions in hadrons. We include charm via the evolution equations
($N_f$), at the appropriate $Q^2$ of charm production, to avoid any
non-empirical assumptions about its size. The entire LO and NLO analysis
is done in $x$-space to avoid the potential pitfalls of losing kinematical
information in the inversion of moments. Physically, the small-$x$ behavior is
of the Regge type, consistent with data and other theoretical approaches
\cite{ek}. The large-$x$ behavior is compatible with the appropriate counting
rules \cite{bbs}.

We consider three distinct models for the polarized gluons, which have a 
moderately wide range. Our choice effectively includes two separate
factorization schemes: Gauge Invariant (GI) and Chiral Invariant (CI). 
These are all physically motivated models, whose overall size not large.
The final parametrizations are easy to use, both in form and format.
They are also in excellent agreement with the most recent data.

\mainhead{II. Theoretical Background}

\subhead{A. Polarized Quark Distributions}

Any distributions that are to be used to predict physical observables must be
consistent with both existing, related data and certain fundamental theoretical
assumptions. In the case of the polarized distributions, the spin information
as it applies to hadronic structure must be implicitly included, and the
appropriate kinematic behavior must be explicit, so that they satisfy the
fundamental constraints. This major requirement covers both the theoretical and
experimental considerations which are important. Thus, we wish to construct
the $x$-dependent polarized valence and sea quark distributions subject to the
following physical constraints:

\begin{itemize}

\item the integration over $x$ should reproduce the values extracted from
PDIS data, so that the fraction of spin carried by each constituent is
contained implicitly in the flavor-dependent distributions

\item the distributions should reproduce the $x$-dependent polarized structure
functions, $g_1^i$, $i$=p,n and d at the average $Q^2$ values of the data

\item the small-$x$ behavior of $g_1(x)$ should fall between a Regge quark-like
power of $x$ and a gluon-dominated logarithmic behavior

\item the $Q^2$ behavior of the quark distributions should be consistent
with the non-singlet and singlet NLO evolution equations, for the number of
flavors appropriate to the $Q^2$ range to be covered

\item the positivity constraints are satisfied for all of the flavors.

\end{itemize}

The first two constraints build in compatibility with both the integrated
and the $x$-dependent polarized deep-inelastic-scattering (PDIS) data. The
third and fourth conditions satisfy sound theoretical assumptions about both
the $x$ and $Q^2$ kinematical dependence of the distributions. Finally, the
last constraint is fundamental to our physical understanding of polarization.

\subhead{Valence and Sea Quark Assumptions}

We construct the polarized valence distributions from the unpolarized
distributions by imposing a modified SU(6) model \cite{ck,qrrs}:
$$
\eqalign{
\Delta u_v(x)&\equiv \cos\theta_D(x) \bigl[u_v(x)-{2\over 3}d_v(x)\bigr], \cr
\Delta d_v(x)&\equiv \cos\theta_D(x) \bigl[-{1\over 3}d_v(x)\bigr],}
\auto\label{2.1}
$$
where the spin dilution factor is given by:
$\cos\theta_D\equiv \bigl[1+R_0 (1-x)^2/\sqrt{x}\bigr]^{-1}$. The $R_0$ term
is chosen to satisfy the Bjorken Sum Rule (BSR), including the appropriate QCD
corrections. In the $Q^2$ region of the present PDIS data, we find that 
$R_0\approx {{2\alpha_s}\over 3}$. We may choose the unpolarized valence
distributions $u_v$ and $d_v$ as either the MRS \cite{mrs}:
$$
\eqalign{
xu_v(x)=2.43 x^{0.6}(1-x)^{3.69}\bigl[1-1.18 \sqrt{x}+6.18x\bigr], \cr
xd_v(x)=0.14 x^{0.24}(1-x)^{4.43}\bigl[1+5.63 \sqrt{x}+25.5x\bigr],}
\auto\label{2.2}
$$
or the CTEQ \cite{cteq}:
$$
\eqalign{
xu_v(x)=1.344 x^{0.501}(1-x)^{3.689}\bigl[1+6.402x^{0.873}\bigr], \cr
xd_v(x)=0.640 x^{0.501}(1-x)^{4.247}\bigl[1+2.690x^{0.333}\bigr],}
\auto\label{2.3}
$$
or equivalent.

If we assume a model of the sea obtaining its polarization from gluon
Bremsstrahlung, then the polarized distributions naively would have a form
$\Delta q_f = xq_f$ \cite{cls,chs}. However, the polarized deep-inelastic
scattering data appear to imply a negatively polarized sea \cite{gr}.
If the integrated structure functions are to agree with data,
then the normalization defined by $\eta_{av}\equiv \VEV {\Delta q}/\VEV {xq}$
must be a part of the proportionality between the polarized and unpolarized
distributions. Meanwhile, the positivity constraint, which requires
$\mid \Delta q\mid \le q$ for all $x$, implies a functional form for the
function $\eta(x)$. The basic idea of the Bremsstrahlung model, coupled with
the implications of the data motivate the following form for the flavor
dependent polarized sea distributions:
$$
\eqalign{
\Delta q_f(x)\equiv \eta_f(x)\>x\>q_f(x),}
\auto\label{2.4}
$$
where $q(x)$ is the $x$-dependent unpolarized distribution for flavor $f$.
The function $\eta(x)$ is chosen to satisfy the normalization constraint:
$$
\eqalign{
\VEV{\Delta q_f}=\int_0^1 \eta_f(x)\>x\>q_f(x)\>dx\equiv \VEV{\eta xq_f}
=\eta_{av} \VEV{xq_f},}
\auto\label{2.5}
$$
where $\eta_{av}$ is extracted from data for each flavor \cite{gr}. Physically,
$\eta(x)$ may be interpreted as a modification of $\Delta q$ due to unknown
effects of soft physics at low $x$. This motivates a form for $\eta(x)$,
which will be discussed later.

\subhead{Positivity Constraint}

For the purposes of this analysis, we are assuming that the sea quarks are
effectively massless, so each quark has a definite helicity state. In essence,
this ignores higher twist transverse spin effects in the entire kinematic
range considered. Thus, there is a probabilistic interpretation of the parton
densities, q, and the net parton distribution is given by:
$q=q\uparrow+q\downarrow$.
The total polarization is given as the difference of probabilities of finding
polarized $\uparrow$ and $\downarrow$ partons in the nucleon:
$\Delta q\equiv q\uparrow-q\downarrow$. In a polarized $\uparrow$ proton,
this probability for sea quarks should be less than that of the total
unpolarized sea distribution, since every quark is in a given helicity state.
This is the positivity constraint; i.e,
$$
\eqalign{
\mid \Delta q(x)\mid \le q(x).}  \auto\label{2.6}
$$
for all $x$.
 
This is valid for the leading order (LO) $x$-dependent distributions which have
a clear probabilistic interpretation, because there is no intrinsic scheme
dependence at this level. The results in the next-to-leading-order (NLO)
treatment depend only upon the GLAP evolution, so there is no problem with
constraining the sea quarks using positivity at $Q_0^2$ in either case.
The valence quarks satisfy the positivity constraint by construction.
Although the probabilistic meaning gives rise to the positivity constraint,
it is not clear what role the chiral and gauge invariant schemes have on
this interpretation. In our treatment, the polarized gluon distribution is
assumed small at these low $Q^2$ values, so the schemes are close enough so that
positivity is unaffected. This provides a motivation to choose a zero $\Delta G$
model - to investigate the gauge invariant factorization.

\subhead{Evolution}

The polarized partons are grouped in a linear combination which is a
singlet of flavor $SU_f(3)$ group and a nonsinglet linear combination of
that group. The singlet/non-singlet designations are useful for delineating
certain evolution and factorization properties of the quarks. These combinations
can be related to the usual flavor decomposition, including the valence, sea
and gluons. The nonsinglet term $\Delta q_{NS}$ is a linear combination 
of the triplet $a_3$ axial charge and an octet $a_8$ axial charge of the
$SU_f(3)$ symmetry. The singlet, $\Delta \Sigma$ is related to the axial
charge, $a_0$ and is normally associated with the axial anomaly, which 
includes a gluonic contribution. This depends upon the factorization scheme,
which will be discussed shortly.

The non-singlet term is dominated by the valence distribution, under polarized
($\Delta u_s$, $\Delta d_s$) symmetry. Its first moment is scale ($Q^2$)
independent in LO. The singlet term has a definite physical interpretation in
the gauge invariant scheme, even though it is scale dependent. Since we are 
going to assume a small $\Delta G$ in all of our models, this interpretation
will be essentially valid, even in the chiral invariant (Adler-Bardeen) scheme,
which separates out the anomaly. Thus, our distributions are constructed within
the positivity constraint, in both cases. In the NLO analysis, the initial 
distributions are constructed at a low enough $Q_0^2$ value to justify the
constraints that we have set. We assume that this $Q_0^2$ is still large enough
that higher-twist effects are negligible \cite{larin}. The distributions are
then evolved in NLO, completely independent of any further considerations. In
the evolution, the separation of singlet and non-singlet is faithfully
maintained. Thus, we can satisfy all of the important physical constraints.

Our parton distributions are evolved directly in $x$-space using an iteration
technique first suggested in reference \cite{rossi} and we use the splitting
functions recently calculated in reference \cite{vogel}. This method differs
from the 'brute force' technique recently used in reference \cite{km} and
requires less computer time than the conventional evolution in $n$-moment
space. In addition it eliminates the need to invert from $n$-moment space and
the attendant difficulties in covering the extreme $x$-values. As a result,
this method is inherently more accurate than evolving the distributions in 
moment space. Details of the technique can be found in reference \cite{goram}.

\subhead{B. The Role of Polarized Gluons}

\subhead{Factorization}

In PDIS, there are two factorization schemes which can be used to represent the 
polarized sea distributions: the gauge-invariant \cite{bq} (or $\overline{MS}$)
and the chiral invariant \cite{ab} (or Adler-Bardeen) schemes \cite{cheng}. In 
the chiral-invariant (AB) scheme, the axial gluon anomaly term, \cite{ga} which
depends upon the polarized gluon distribution, is separated out from the 
chiral invariant polarized quark distributions. Since the measured 
distributions must be gauge invariant, the relation between the two scheme
dependent distributions is:
$$
\eqalign{
\Delta q(x)_{GI}=\Delta q(x)_{CI}-{{N_f\alpha_s}\over {2\pi}}\Delta G(x),}
\auto\label{2.6a}
$$
where the GI refers to the gauge-invariant scheme and CI to the chiral-invariant
scheme. Thus, the size of $\Delta G$ is relevant in the CI scheme.
This, in turn, affects the average $\eta$ values extracted for each flavor.
Thus, $\Delta G$ has an indirect bearing on the polarized sea distributions.
\cite{gr} 

There exists no empirical evidence that the polarized gluon distribution is
very large at the relatively small $Q^2$ values of the data. In fact, even in
the NLO evolution, the polarized gluon distribution does not evolve
significantly between $Q^2=1$ and $Q^2=10$ GeV$^2$, regardless
of which model is chosen. Data from Fermilab \cite{e704} indicate that it is
likely small at the $Q^2$ values of existing data. In addition, a theoretical
model of the polarized glue, based on counting rules, implies that 
$\Delta G\approx {1\over 2}$ \cite{bbs}. Other theoretical models substantiate
this claim, as well \cite{bcd,bj}. However, since we do not know the
explicit size of the $\Delta G$ in this kinematic region, we perform our 
analysis using three distinct physical models. The evolution of $\Delta G$ is
then done both in LO and NLO to investigate any possible NLO effects.

The first set of $\eta(x)$ functions, quoted in Table I assumes a moderately
polarized glue:
$$
\eqalign{
\Delta G(x)=xG(x)=31.3 x^{0.41}(1-x)^{6.54}\bigl[1-4.64 \sqrt{x}+6.55x\bigr],}
\auto\label{2.12}
$$
using an unpolarized MRS glue, normalized to 0.50 or
$$
\eqalign{
\Delta G(x)=xG(x)=1.123 x^{-0.206}(1-x)^{4.673}\bigl[1+4.269x^{1.508}\bigr],}
\auto\label{2.13}
$$
for the CTEQ gluon distribution, consistent with the analysis in reference 1. 

For the second polarized gluon model, we set $\Delta G=0$ to determine
$\eta_{av}$ and parametrize the $\eta(x)$ accordingly. This is equivalent
to the gauge-invariant scheme, since the anomaly term vanishes. Any analysis
which requires a gauge-independent set of flavor dependent distributions (such
as on the lattice) should use this set of polarized distributions at $Q_0^2$,
evolved to the appropriate $Q^2$ values. The corresponding $\eta(x)$ functions
are listed in Table II.

The third gluon model is motivated by an instanton-induced polarized
gluon distribution, which gives a negatively polarized glue at small-$x$
\cite{koch}. This modified distribution (normalized to $\Delta G=-0.23$) is
given by the best fit to the curve in reference \cite{koch}:
$$
\eqalign{
\Delta G(x)=7 (1-x)^7\bigl[1+0.474\ln(x)\bigr].}
\auto\label{2.14}
$$
This would allow for instanton based non-perturbative effects at small $Q^2$.

\subhead{Relation between the sea and glue in $g_1$}

The valence polarizations are rather well established from the BSR and
the polarized DIS experiments. The valence quarks are dominant at large $x$
and they give their polarization to the gluons through Bremsstrahlung, which
in turn, creates sea polarization via pair-production. But the sea quarks 
share the momentum from the gluon which created the pair, and thus, each 
constituent is at lower $x$ than the original valence "parent". This is 
consistent with the polarized sea being dominant at lower $x$. The PDIS data
imply that the sea is polarized opposite to that of the valence quarks.
The relative size of the negative sea polarization is an indication as to
whether gluon polarization is moderately positive (such as $\Delta G =xG$,
implying that polarization of G carries most of the spin of the proton), nil, or
negative. We expect a larger negatively polarized sea for the last two cases
as it must offset the positive anomaly term proportional to gluons in the
chiral-invariant scheme and the $J_z={1\over 2}$ sum rule in either scheme.
If the polarized sea is smaller (less negative) or the polarized gluon is 
large, the $g_1^p$ curve will exhibit a sharper rise at small $x$. When future
data are available with smaller error bars, these scenarios can be better
defined to yield the correct sign and size of $\Delta G$.
There may be a possibility to argue a positive proportionality of spin and
momentum of the sea if $\Delta G$ is very large. A much larger polarized gluon
distribution can imply either a smaller negatively polarized sea or even a
slightly positive polarized sea. However, this would require a prohibitively
large polarized glue at these smaller $Q^2$ values. We argue that this is not
likely for the following reasons:

\begin{itemize}
\item (1) when the light-cone wavefunctions at small-$x$ are analyzed, they
implicate a negatively polarized sea, \cite{abd}
\item (2) in order for the sea to be entirely positive, $\Delta G$ would
have to be prohibitively large to satisfy the data. The orbital angular
momentum would have to be correspondingly large to satisfy the J=1/2 sum rule.
This would also disagree with the implications of the E704 data, \cite{e704}
\item (3) Our highly successful fits to the $x$-dependent data not only indicate
that $\Delta G$ is likely moderate at $Q^2$=10 GeV$^2$, but that the data at
lower $Q^2$ is fit somewhat better with even smaller $\Delta G$. This is
consistent with the evolution of the polarized gluon distribution. We also show
in Section IV that the growth of $\Delta G$ presented by ABFR \cite{bfr} is not
likely, even in NLO. Thus, our present analysis further strengthens the
point of a smaller $\Delta G$ at these lower $Q^2$ values,
\item (4) most all other independent analyses agree with the negatively
polarized sea.
\end{itemize}

We conclude that a positively polarized sea and a very large $\Delta G$ seem
unlikely, given present data.

\subhead{C. Extrapolation of Data to Small-x}

Extrapolation of $g_1^i$ ($i=p,n,d$) to small-$x$ is important experimentally
for determination of the integrated values of these structure functions.
Theoretically, the $g_1$ behavior at small-$x$ is important to understanding the
mechanisms which underlie the physics in this region. There are various models
that attempt to explain the contributions to both $F_2$ and $g_1$ at low-$x$.
The data appear to exhibit growth of these quantities, but since the error bars
are somewhat large, most of the predicted types of behavior cannot be ruled
out \cite{cr}.

At large-$x$, the valence distributions dominate $F_2$ and $g_1$ and these
are more well determined than the sea distributions, which are more prevalent at
small-$x$. Thus, one must make suitable assumptions about the behavior of the
polarized sea at low-$x$. Experimental analyses \cite{smc,e142} have tended to
assume a relatively constant behavior and extrapolate $g_1$ from its value at
about $x\sim 10^{-2}$ down to $x=0$. A model by Donnachie and Landshoff
\cite{dl} assumes that the Pomeron couples via vector $\gamma_{\mu}$ so that
$g_1$ exhibits a logarithmic behavior: $g_1\sim \ln(1/x)$. Bass and Landshoff
\cite{bl} analyze a model of a two-gluon Pomeron which leads to a slightly more
divergent behavior at small-$x$: $g_1\sim \bigl[1+2\ln(x)\bigr]$. If negative
parity Pomeron cuts contribute to the spin-dependent cross section, a divergent
behavior of $g_1$ results, \cite{sing} corresponding to the singular form:
$g_1\sim 1/x\ln^2(x)$.

We make no presumptions about the forms of the flavor dependent distributions
at small-$x$, other than their relation to the unpolarized distributions.
The parametrization of $\eta_f(x)$ in (2.4) will be determined primarily from
normalization and positivity constraints. Once the polarized sea flavors are
generated, we can determine resulting the small-$x$ behavior of $g_1$ and
compare it to these theoretical models.

\mainhead{III. Phenomenology}

\subhead{A. Polarized Quark Flavors}

In the work of reference 1, the integrated polarized structure functions
were compared to the spin averaged distributions, to establish a comparison
between the spin and momentum carried by each flavor of quark. The results
indicate that the relation is flavor dependent, but the magnitudes of these
ratios ($\eta_{av}$ in equation 2.5) are of the same order of magnitude. 
Although this does not necessarily imply that there is a direct relation
between the two, it does provide a suitable starting point for generating the
polarized distributions from known unpolarized distributions, which satisfy
the data on spin-averaged physical processes.
This is the motivation for the form in equation (2.4) for the polarized
flavor-dependent distributions. The integrated data are satisfied by choosing
the parameters in $\eta(x)$ to satisfy equation (2.5) and the positivity
constraints. We also wish to stay consistent with counting rules at large-$x$
\cite{bbs}. The small-$x$ behavior can be controlled by the functional form that
we choose for $\eta(x)$. All of these constraints are to be satisfied at some
low value of $Q^2$, and the evolution equations will ensure that positivity
and the kinematical behavior stay consistent at all $Q^2$.

A possible form for $\eta(x)$, which gives flexibility in satisfying
the constraints (2.5) and (2.6) is: $\eta(x)=a+bx^n$.
We expect the function to be decreasing with $x$, since the problems with
positivity (for $\mid \eta_{av}\mid > 1$) occur at large $x$. We chose not
to modify the $(1-x)$ dependence in order to keep the counting rule powers
in tact (insofar as the unpolarized distributions do this) for the large-$x$
behavior. We were able to satisfy the positivity constraint at all $x$ using
this form for $\eta(x)$.

In the following analysis, we assume the unpolarized distributions in the
CTEQ form: $q(x)=A_0x^{A_1}(1-x)^{A_2}(1+A_3 x^{A_4})$. Then, using equation
(2.4), we generate the corresponding polarized distributions for each flavor. 
An analysis with the MRS distributions yields similar results.

For the CTEQ distributions, we have
$$
\eqalign{
x\bar{q}(x)={1\over 2}\Bigl[0.255 x^{-0.143}(1-x)^{8.041}(1+6.112x)\mp
0.071 x^{0.501}(1-x)^{8.041}\Bigr],}
\auto\label{3.3}
$$
where the ($-$) holds for $\bar{u}$ and the (+) for the $\bar{d}$ flavors.
The strange sea has the parametrization:
$$
\eqalign{
x\bar{s}(x)=\Bigl[0.064 x^{-0.143}(1-x)^{8.041}(1+6.112x)\Bigr].}
\auto\label{3.4}
$$
Both of these sets account for the $\bar{u}$, $\bar{d}$ asymmetry in the
unpolarized sea.
 
The corresponding integrated polarized distributions can be written in terms
of beta functions, $B(m,n)$, as:
$$
\eqalign{
\VEV{\Delta q}&=aA_0 B(A_1+2,A_2+1)+aA_0 A_3 B(A_1+A_4+2,A_2+1) \cr
&-bA_0 B(A_1+n+2,A_2+1)-bA_0 A_3 B(A_1+A_4+n+2,A_2+1)}  \auto\label{3.6}
$$
for the CTEQ distributions.
The integral $\VEV{xq}$ can be similarly written and thus the restriction
on $a$ and $b$, corresponding to the normalization constraint (2.5) is:
$a=\eta_{av}-b\lambda$, where
$$
\eqalign{
\lambda\equiv {{B(A_1+n+2,A_2+1)+A_3 B(A_1+A_4+n+2,A_2+1)}\over {B(A_1+2,A_2+1)
+A_3 B(A_1+A_4+2,A_2+1)}},}  \auto\label{3.8}
$$
for the CTEQ unpolarized distributions.

The positivity constraint in terms of $a$ and $b$ is
$$
\eqalign{
{{\mid \Delta q(x)\mid}\over {q(x)}}=\mid ax+bx^{n+1}\mid \le 1,}
\auto\label{3.9}
$$
for all $x\in [0,1]$. In the following discussion, we will show how $\eta(x)$
is generated for the zero polarized gluon case. The procedure is virtually
identical to the other two gluon models where the anomaly term is present.

When we choose $a$ to satisfy the normalization constraint (2.5) and $b$ to
satisfy the positivity constraint (2.6), the $x$-dependent polarized
distribution for each flavor is determined from (2.4). These conditions are
independent of the set of unpolarized distributions that is used. However,
one must be consistent by using the appropriate set of unpolarized
distributions which were used to determine the function $\eta(x)$.
Therefore, we seek the form 
$$
\eqalign{
\eta(x)=(\eta_{av}-b\lambda)+bx^n,}  \auto\label{3.10}
$$
where both $b$ and $n$ are chosen to satisfy the positivity constraint (3.5).
This form of $\eta(x)$ is motivated by: (1) associating this function with
modifications of $\Delta q$ by small-$x$ physics and (2) keeping the $(1-x)$
dependence in tact to be consistent with counting rule behavior at large $x$
to the extent that the unpolarized distributions have this desired form
\cite{bbs}. We can then both satisfy positivity (even at large $x$) and 
control the small $x$ behavior, where the sea and glue are most prominent.

\subhead{Choice of $\eta(x)$ at small-$x$}

The polarized sea and the gluon distributions are expected to dominate in the
small-x region. If we assume a strongly polarized negative sea, with
$\eta_{av} <-1$ then this defines a range of $b$ values which satisfy the
positivity constraint. The behavior of our Ansatz $\eta(x)=a+bx^n$ for
$0<n<1$ is suited for small-x dependence. When a certain small-$x$ behavior is
desired, a series of n values can be tried for the best fits to data. In fact,
when $0.2<n<1.0$, it is easier to satisfy both constraints with appropriate 
choices of $a$ and $b$. These values of $n$ allow a wider range of $a$ and $b$ 
values. However, when $0<n<0.2$, the range of possible $a$ and $b$ values
which satisfy the constraints gets very small and will not be the same for all
experimental yields of $\eta_{av}$. Thus, it makes it impossible to find a
uniform fit for $\eta(x)$ for each flavor. Since $\eta(x)$ should be only
flavor dependent to have any physical connection, we must choose $n$ so that
$a$ and $b$ will be comparable for all experimental results.

Positivity of $n$ is, in principle, not essential. If we choose a small negative
$n$, we enhance the divergence of $g_1$ as $x\to 0$. This favors strong 
anti-polarization of sea at small-$x$ and could give the sea some positive
polarization for large $x$. However, it is virtually impossible to satisfy both
the normalization and positivity constraints simultaneously with negative $n$
and such a simple parametrization of $\eta$. Thus, it does not appear to
be advantageous to choose $n$ to be negative, especially since we have fit the
data successfully with $n={1\over 2}$.

The ranges for possible $\eta(x)$ functions are given in Tables I through III.
The corresponding polarized distributions satisfy the DIS data and the
positivity constraint. 

\subhead{B. Input from DIS Data}

In Tables I through III, we summarize the integrated results for each
considered experiment, with the values for $\eta_{av}$ in each gluon model.
The corresponding functional form for each $\eta(x)$ is shown, which satisfies
the constraints discussed in the text. We chose two sets of data each, for the
proton, \cite{e143,smc} neutron \cite{e154,hermes} and deuteron \cite{e143,
smc}. These represent the latest published data and are representative of
the groups at SLAC, CERN and DESY.

\tablehead{Table I: $\eta$ Values from Data and $\eta(x)$: $\Delta G=xG$}
$$\begin{array}{ccccc}
 Quantity    & \eta_{u,d} & \eta_s & \eta_u(x)       &\eta_s(x)  \cr
 SMC\>(I^p)    & -2.0   & -1.6   & -2.84+2.8\sqrt x  & -2.23+2.1\sqrt x \cr
 E143\>(I^p)   & -1.8   & -1.2   & -2.64+2.8\sqrt x  & -1.83+2.1\sqrt x \cr
 E154\>(I^n)   & -1.5   & -0.6   & -2.34+2.8\sqrt x  & -1.23+2.1\sqrt x \cr
 HERMES\>(I^n) & -1.3   & -0.3   & -2.14+2.8\sqrt x  & -0.93+2.1\sqrt x \cr
 E143\>(I^d)   & -1.6   & -0.8   & -2.44+2.8\sqrt x  & -1.43+2.1\sqrt x \cr
 SMC\>(I^d)    & -2.4   & -2.3   & -3.24+2.8\sqrt x  & -2.93+2.1\sqrt x 
\end{array}$$

\tablehead{Table II: $\eta$ Values from Data and $\eta(x)$: $\Delta G=0$}
$$\begin{array}{ccccc}
 Quantity    & \eta_{u,d} & \eta_s & \eta_u(x)   &\eta_s(x)  \cr
 SMC\>(I^p)    & -2.4   & -2.2   & -3.30+3.0\sqrt x  & -3.09+2.9\sqrt x \cr
 E143\>(I^p)   & -2.2   & -2.0   & -3.10+3.0\sqrt x  & -2.87+2.9\sqrt x \cr
 E154\>(I^n)   & -1.8   & -1.3   & -2.70+3.0\sqrt x  & -2.17+2.9\sqrt x \cr
 HERMES\>(I^n) & -1.7   & -1.0   & -2.60+3.0\sqrt x  & -1.90+2.9\sqrt x \cr
 E143\>(I^d)   & -2.0   & -1.6   & -2.90+3.0\sqrt x  & -2.47+2.9\sqrt x \cr
 SMC\>(I^d)    & -2.7   & -2.9   & -3.60+3.0\sqrt x  & -3.77+2.9\sqrt x 
\end{array}$$

\tablehead{Table III: $\eta$ Values from Data and $\eta(x)$: Neg. $\Delta G$}
$$\begin{array}{ccccc}
 Quantity    & \eta_{u,d} & \eta_s & \eta_u(x)   &\eta_s(x)  \cr
 SMC\>(I^p)    & -2.5   & -2.5   & -3.43+3.1\sqrt x  & -3.49+3.3\sqrt x \cr
 E143\>(I^p)   & -2.4   & -2.4   & -3.33+3.1\sqrt x  & -3.39+3.3\sqrt x \cr
 E154\>(I^n)   & -2.0   & -1.7   & -2.93+3.1\sqrt x  & -2.99+3.3\sqrt x \cr
 HERMES\>(I^n) & -1.9   & -1.5   & -2.83+3.1\sqrt x  & -2.89+3.3\sqrt x \cr
 E143\>(I^d)   & -2.2   & -2.0   & -3.13+3.1\sqrt x  & -3.19+3.3\sqrt x \cr
 SMC\>(I^d)    & -2.9   & -3.2   & -3.83+3.1\sqrt x  & -3.89+3.3\sqrt x 
\end{array}$$

Since our ultimate goal is to find a suitable set of flavor-dependent $\eta (x)$
functions, which do not depend upon a specific experimental result, we take a
suitable average of $\eta(x)$ for each flavor to generate the polarized
sea quark distributions. There is enough flexibility in the choice of
$a$ and $b$, given the experimental errors and the range of values which
satisfy positivity, so that all constraints are still satisfied. Note from
the tables that the range of $a$ values is not considerable, even when the
values of $b$ are fixed. Our choice of the averaging procedure is further
justified by our ability to reproduce the data from all of the experiments.
The resulting functions $\eta (x)$ for each gluon model are:
$$\begin{array}{ccc}
 Quantity     & \eta_{u,d}(x)     & \eta_s(x)  \cr
 \Delta G=xG  & -2.49+2.8\sqrt x  & -1.67+2.1\sqrt x \cr
 \Delta G=0   & -3.03+3.0\sqrt x  & -2.71+2.9\sqrt x \cr
 \Delta G<0   & -3.25+3.1\sqrt x  & -3.31+3.3\sqrt x
\end{array}$$

\mainhead{IV. Results and Discussion}

\subhead{A. Results for the Polarized Distributions}

The polarized valence quark distributions are constructed with the assumptions
made in eqns. (2.1) and (2.3), with $R_0$ determined by the BSR.
The overall parametrization for each of the polarized sea flavors, including
the $\eta(x)$ functions, the anomaly terms and the up-down unpolarized
asymmetry term can be written (with the CTEQ basis) in the form:
$$
\eqalign{
\Delta q_i(x)=-Ax^{-0.143}(1-x)^{8.041}(1-B\sqrt{x})\Bigl[1+6.112x+P(x)
\Bigr].}   \auto\label{4.1}
$$
The values for the variables in equation 4.1 are given for each flavor and
each gluon model in Table IV.

\tablehead{Table IV: Parametrizations for Polarized Sea Flavors}
$$\begin{array}{|c||c||c|c|c|}
\hline
  Flavor    &\Delta G  	&A  	&B  	&P(x) 			\cr
\hline                                                  
\hline
  <\Delta u>_{sea} &xG 	&0.317 &1.124 &-0.278x^{0.644}-1.682x^{0.937}
(1-x)^{-3.368}(1+4.269x^{1.508})	\cr
\hline
  <\Delta d>_{sea} &xG 	&0.317 &1.124 &+0.278x^{0.644}-1.682x^{0.937}
(1-x)^{-3.368}(1+4.269x^{1.508})	\cr
\hline
  <\Delta s> 	&xG 	&0.107 &1.257 &-3.351x^{0.937}(1-x)^{-3.368}
(1+4.269x^{1.508})	\cr
\hline
  <\Delta u>_{sea} &0 	&0.386 &0.990 &-0.278x^{0.644}	\cr
\hline
  <\Delta d>_{sea} &0 	&0.386 &0.990 &+0.278x^{0.644}	\cr
\hline
  <\Delta s> 	&0 	&0.173 &1.070 &0			\cr
\hline
  <\Delta u>_{sea} &Neg &0.414 &0.954 &-0.278x^{0.644}-10.49x^{1.143}
(1-x)^{-1.041}(1+0.474\ln{x})	\cr
\hline
  <\Delta d>_{sea} &Neg &0.414 &0.954 &+0.278x^{0.644}-10.49x^{1.143}
(1-x)^{-1.041}(1+0.474\ln{x})	\cr
\hline
  <\Delta s> 	&Neg 	&0.212 &0.997 &-20.89x^{1.143}(1-x)^{-1.041}
(1+0.474\ln{x})			\cr
\hline
\end{array}$$ 

We have used these to calculate the polarized structure functions, $xg_1(x)$,
for the proton, neutron and deuteron. These are all compared with the
corresponding data at the average $Q^2$ value for that data set. These plots
are shown in figures 1 through 4. In these figures, the solid line corresponds
to the small polarized gluon model, the dashed line to the zero polarized glue
and the dotted lines to the instanton motivated gluon model. 
\begin{figure}
{\hskip 5.0cm}\hbox{\epsfxsize=9.0cm\epsffile{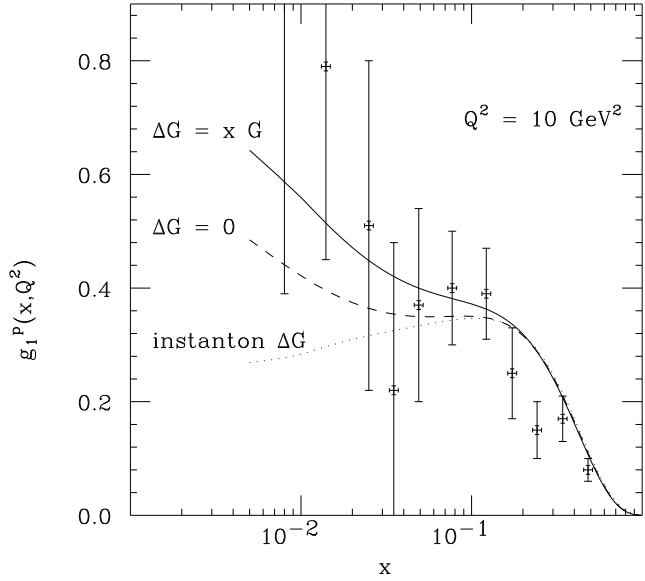}}
\caption{The polarized proton structure $g^p_1$ as a function of $x$ at
fixed $Q^2$ for three models of $\Delta G$ compared to data, and
highlighting small $x$ behaviour.}
\end{figure}
\begin{figure}
{\hskip 5.0cm}\hbox{\epsfxsize=9.0cm\epsffile{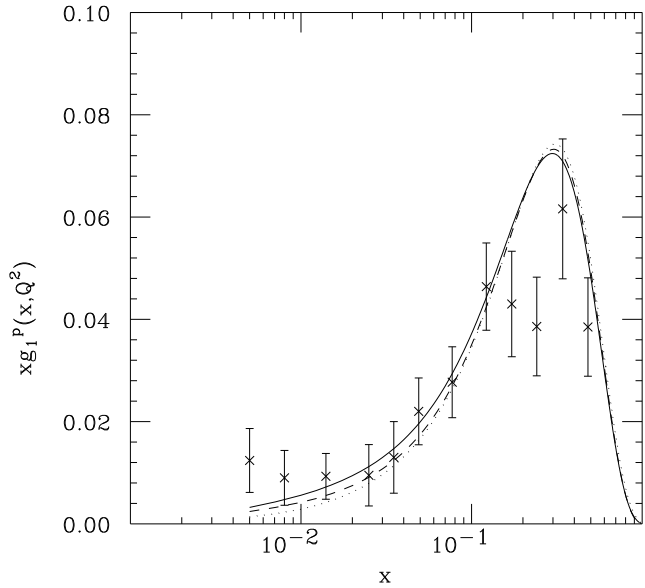}}
{\vskip 2.0cm}\caption{Same as fig.1 but for $xg^p_1$ highlighting
medium $x$ behaviour.}
\end{figure}

\begin{figure}
{\hskip 5.0cm}\hbox{\epsfxsize=9.0cm\epsffile{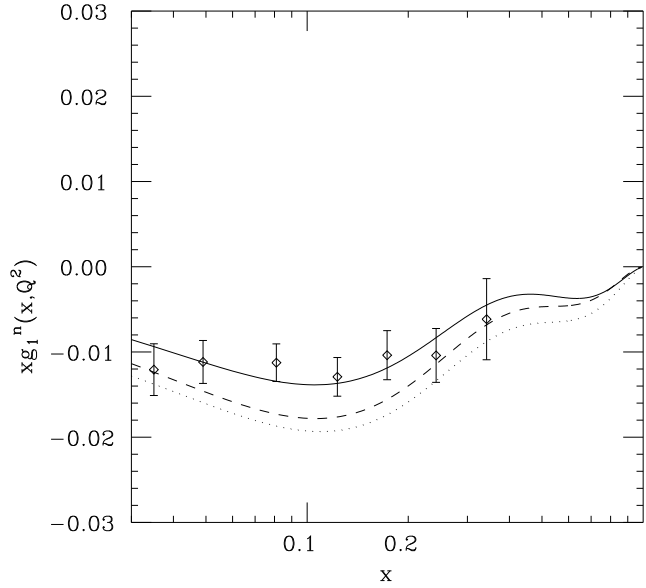}}
{\vskip 1.0cm}\caption{The Polarized Neutron Distribution $xg^n_1$ as a
function of $x$ at $Q^2=10 GeV^2$ compared to data. The three curves are
for three different gluon models (see text).}
\end{figure}
\begin{figure}
{\hskip 5.0cm}\hbox{\epsfxsize=9.0cm\epsffile{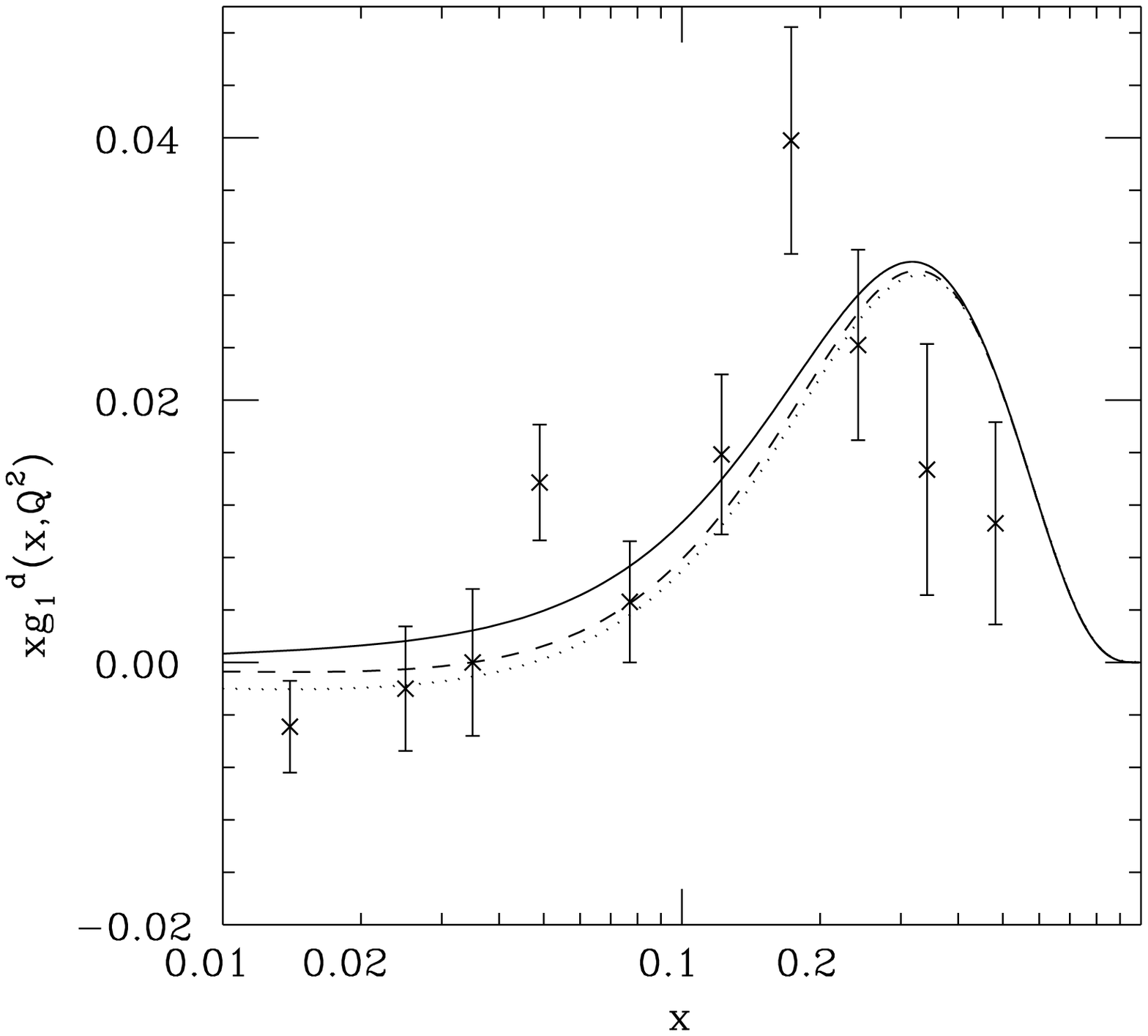}}
{\vskip 1.0cm}\caption{Same as fig.3 for $g^d_1$.}
\end{figure}
In order to verify that our models for $\Delta G$ were reasonable, considering
that the evolution governs the $Q^2$ behavior of the distributions, we evolved
$\Delta G(x,Q_0^2=1)$ to $Q^2=1000$ GeV$^2$ for each model using both LO and
NLO singlet evolution. The $x$-behavior of the gluon distributions is shown
in figures 5 through 7 at the appropriate orders of magnitude of $Q^2$.
The corresponding integrated values for these evolved distributions are shown
in the tables below. 

\tablehead{Leading order polarized gluon evolution: $\int_{x_{min}}^1 
\Delta G\>dx$}

$$\begin{array}{cccc}
 Q^2 (GeV^2)	& \Delta G=xG 	& \Delta G=0	& Instanton 	\cr
 1  		& 0.387  	& 0.071		& -0.076	\cr
 10  		& 0.651  	& 0.107		& +0.045	\cr
 100  		& 0.736  	& 0.167		& +0.118	\cr
 1000  		& 0.794  	& 0.211		& +0.182	\cr
\end{array}$$

\tablehead{Next-to-leading order polarized gluon evolution: 
$\int_{x_{min}}^1 \Delta G\>dx$}

$$\begin{array}{cccc}
 Q^2 (GeV^2)	& \Delta G=xG 	& \Delta G=0	& Instanton 	\cr
 1  		& 0.424  	& 0.080		& -0.082	\cr
 10  		& 0.653  	& 0.119		& +0.047	\cr
 100  		& 0.751  	& 0.183		& +0.130	\cr
 1000  		& 0.811  	& 0.229		& +0.190	\cr
\end{array}$$

\begin{figure}
{\hskip 5.0cm}\hbox{\epsfxsize=9.0cm\epsffile{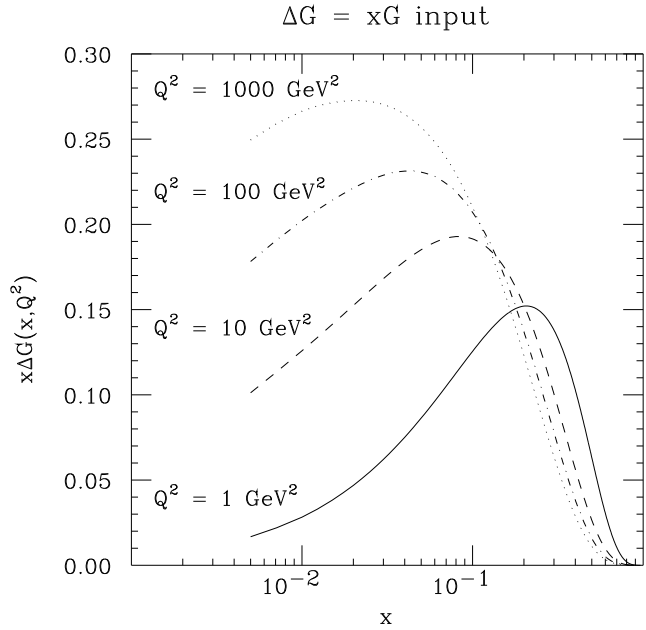}}
{\vskip 1.0cm}\caption{Polarized Gluon Distribution as a function of $x$ 
at different  $Q^2$ values for the $\Delta G = xG$ input.}
\end{figure}
\begin{figure}
{\hskip 5.0cm}\hbox{\epsfxsize=9.0cm\epsffile{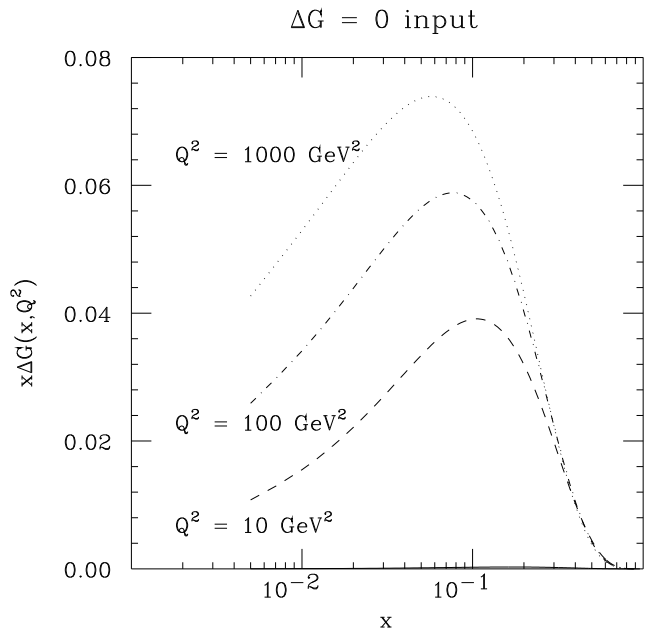}}
{\vskip 1.0cm}\caption{Same as fig.5 for the $\Delta G = 0$ input.}
\end{figure}

\begin{figure}
{\hskip 5.0cm}\hbox{\epsfxsize=9.0cm\epsffile{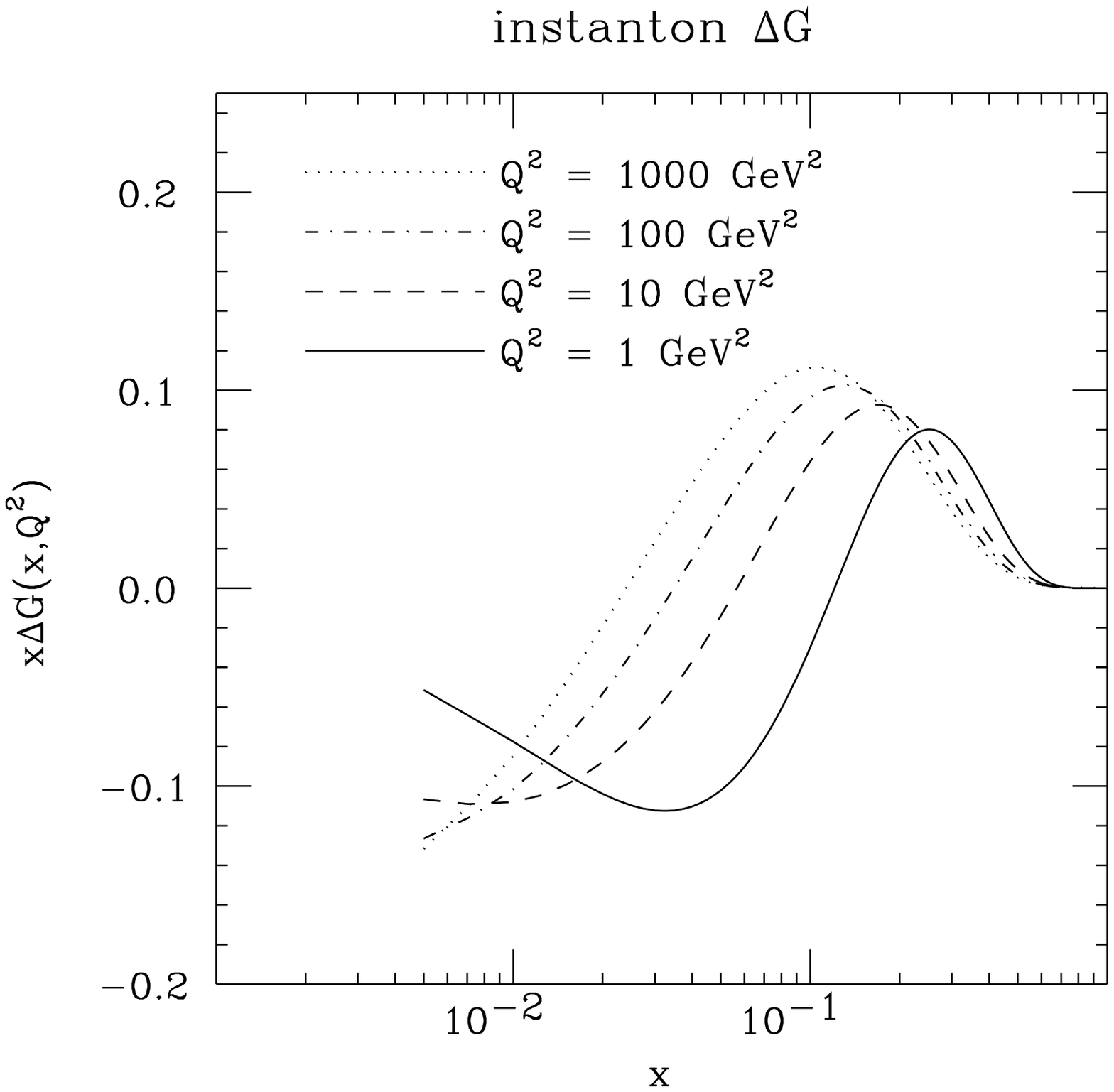}}
{\vskip 5.0cm}\caption{Same as fig.5 and fig.6 for the Instanton gluon
input.}
\end{figure}

For comparison with other models of the polarized quarks, we show the
$x$-dependent distributions of the valence and sea for each flavor in figures
8-11. The sea flavors are shown for each gluon model. Note that our results
compare favorably with other models. There seems to be a general agreement
about the shape of these distributions. Differences arise in the actual
numerical values of the integrated distributions. Both our $x$-dependent and
our integrated distributions have been constructed to satisfy all of the present
data.

\begin{figure}
{\hskip 5.0cm}\hbox{\epsfxsize=9.0cm\epsffile{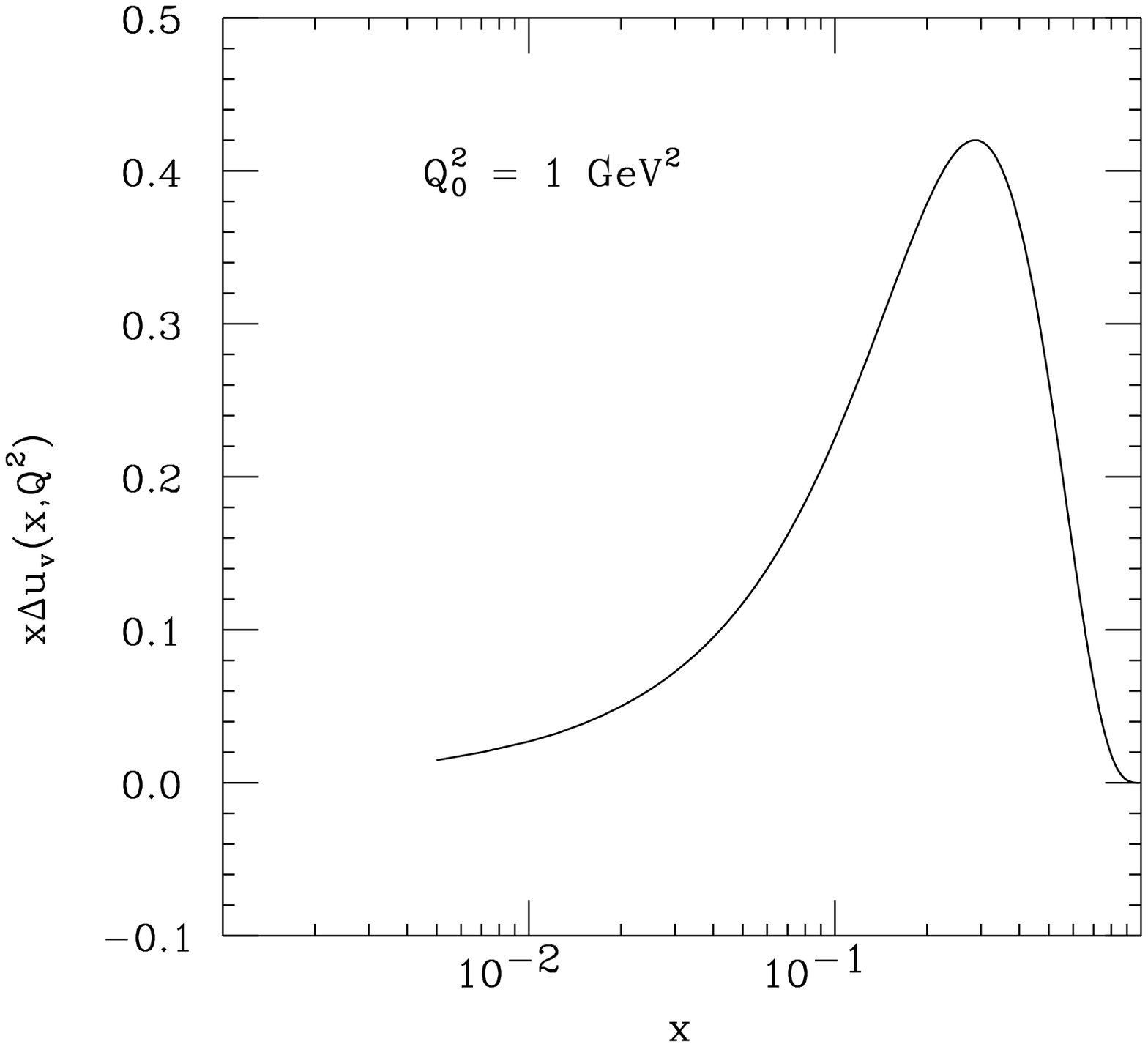}}
{\vskip 1.0cm}\caption{Polarized Valence up quark ($u_v$) Distribution
at low $Q^2$ for the different gluon models.}
\end{figure}
\begin{figure}
{\hskip 5.0cm}\hbox{\epsfxsize=9.0cm\epsffile{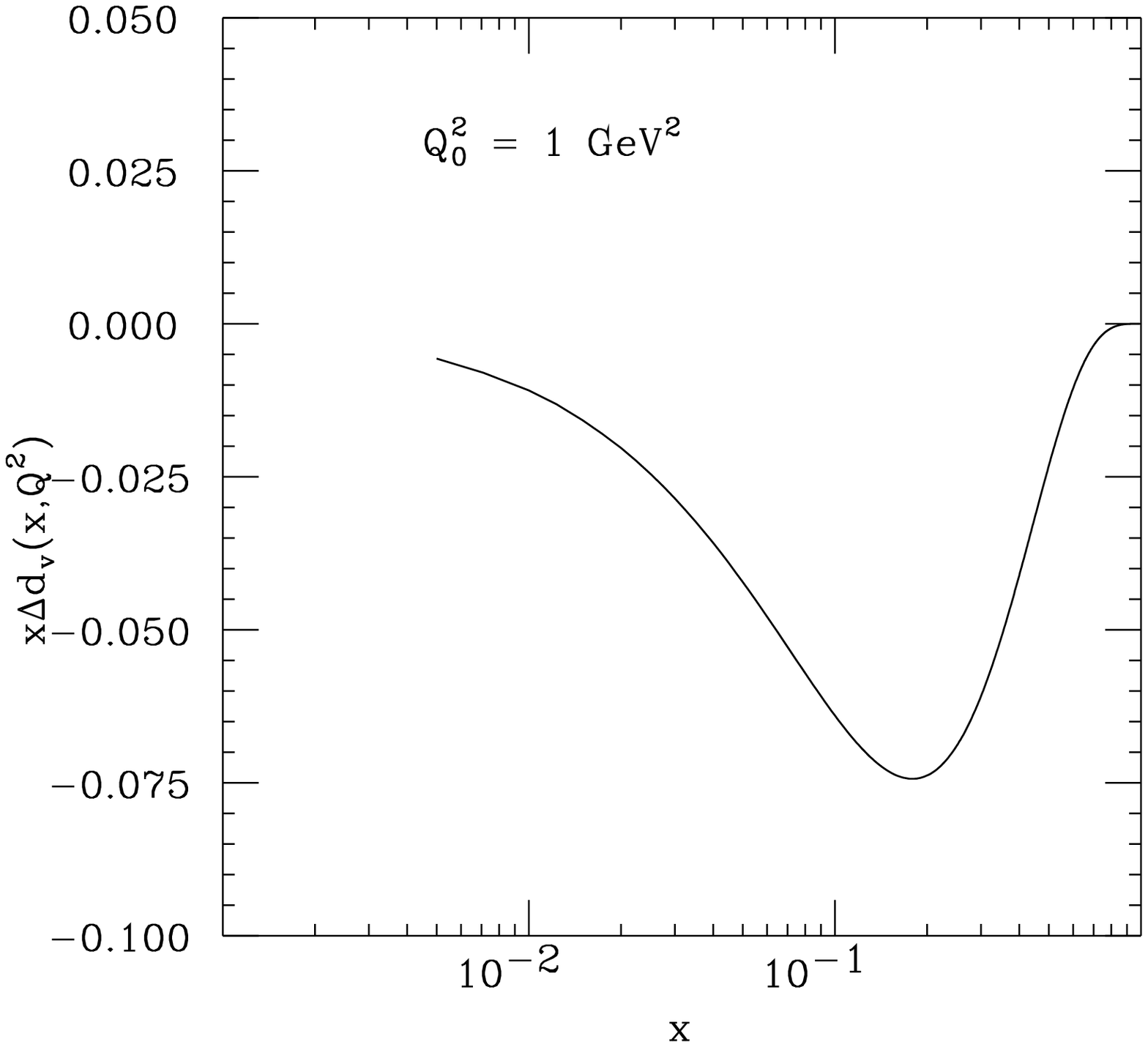}}
{\vskip 1.0cm}\caption{Polarized Valence down quark ($d_v$) Distribution
at low $Q^2$ for the different gluon models.}
\end{figure}

\begin{figure}
{\hskip 5.0cm}\hbox{\epsfxsize=9.0cm\epsffile{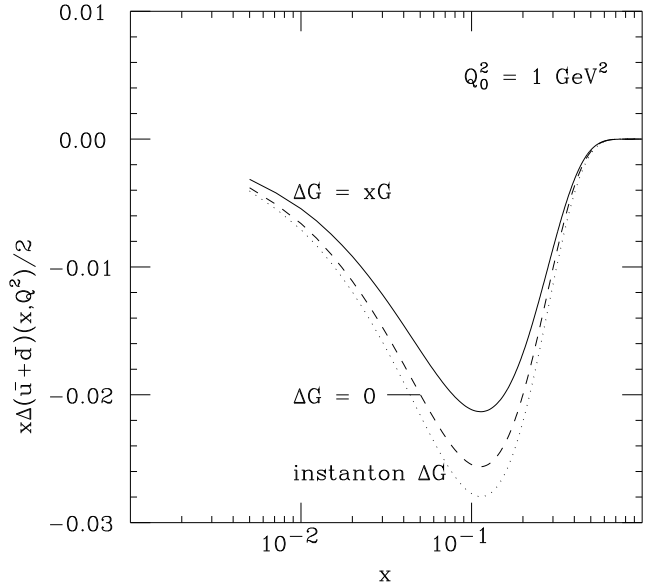}}
{\vskip 1.0cm}\caption{Polarized Up and Down Sea ($\bar{u}+\bar{d}$) 
Distribution at low $Q^2$ for the different gluon models.}
\end{figure}
\begin{figure}
{\hskip 5.0cm}\hbox{\epsfxsize=9.0cm\epsffile{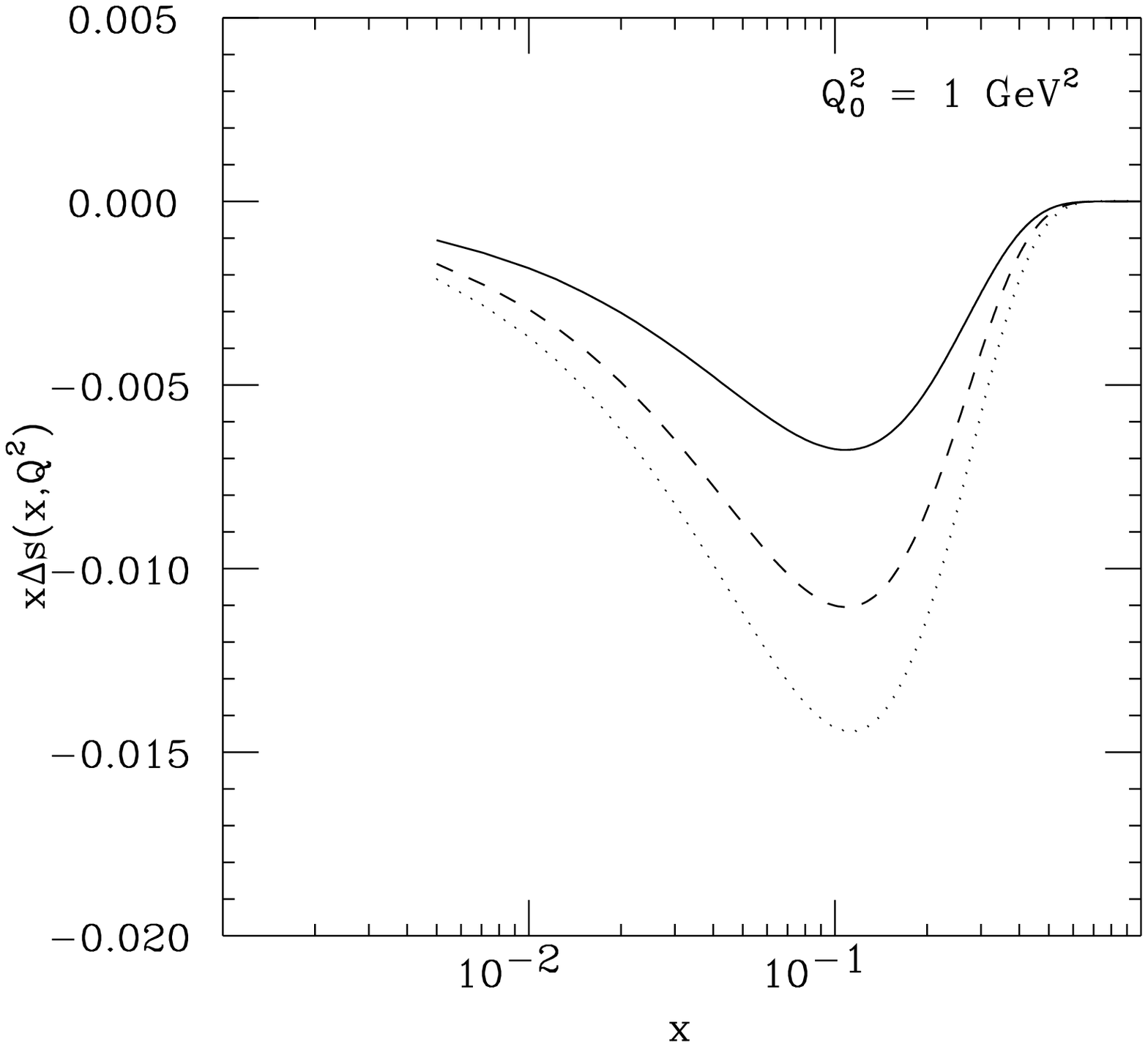}}
{\vskip 1.0cm}\caption{Polarized Strange Sea ($\bar{s}$) Distribution at
low $Q^2$ for the different gluon models.}
\end{figure}

\subhead{Physics Implications}

1. All comparisons of our distributions with existing data are excellent,
including Fig. 1, which shows $g_1$, as opposed to $xg_1$, accentuating the
small-$x$ behavior. The best overall fits occur with the moderate glue
(model 1). The zero glue model results are somewhat better for the neutron,
where the data are at lower average $Q^2$ values. This is consistent with the
$Q^2$ evolution of the polarized gluon distribution.

2. At small-$x$, the instanton gluon model predicts that $g_1^p$ decreases
slightly. However, considering the latitude in this distribution, it is
consistent with a constant behavior. The data appear to be rising in this $x$
region, contrary to this implication. Since the data are at average $Q^2$
of 10 GeV$^2$, this seems to indicate that the gluons are not negatively
polarized at such a relatively large $Q^2$. This is consistent with the
assumption that instantons are dominant at smaller $x$ and $Q^2$ values and
are likely not a major contributor to the polarized glue at higher
$Q^2$ \cite{koch}.

3. The polarized gluon distribution does not evolve as large as BFR predict,
\cite{bfr} even with the moderate gluon model. Assumption of such large
polarization at these lower $Q^2$ values is unfounded. In fact, data from E704
at Fermilab indicate that it is likely more on the order of the moderate or
zero distribution. Even the NLO integrated polarized gluons do not evolve
significantly different from the LO distributions.

4. Our up and down valence distributions are comparable to others. Ours is
motivated from the physical SU(6) model with the BSR fixing the lone free
parameter. It is compatible with the u-valence domination at large $x$ and
has the appropriate $x$-dependent behavior at all other $x$ values.

5. The u and d polarized sea distributions are not highly dependent on the
gluon model used to generate them. However, the polarized strange sea is
quite sensitive to the gluon model and hence the anomaly term. This is
discussed in more detail in reference \cite{gr}.

6. The shape of the $x$-dependent polarized sea distributions agrees with
the analysis of Antonuccio, \etal \cite{abd}. They exhibit Regge-like behavior
at small-$x$ and become slightly positive at moderate $x$. Although it is not
completely obvious from the figures, our sea distributions remain negative
until about $x\sim 0.3$ and then turn slightly positive. This is hidden by the
dominance of the valence quarks in this kinematic range, but indicates a
consistency with physical expectations of the polarized sea. 

\subhead{B. Small-$x$ Behavior}

For the SMC proton data with the CTEQ unpolarized distributions and the
positive gluon model, we find at small $x$ that: $g_1^p\sim x^{-0.19}$.
Phenomenologically, this is due to the interplay between the sea distributions,
with a $\Delta q_i\sim x^{-0.143}$ behavior in eqn. (3.1) and the gluons in 
the model, dominated by $xG\sim x^{-0.206}$ at small-$x$ in eqn. (2.9). 
Physically, this is consistent with Regge behavior, characteristic of the 
iso-triplet contributions to $g_1$. It does not have the steep rise 
characteristic of the singlet behavior due to gluon exchange, but is slightly
steeper than the quoted Regge intercept \cite{heim}. This could either be due
to the uncertainty in the value of the Regge intercept \cite{ek2} or to an
interplay between the quarks and the logarithmic gluon exchange \cite{bl}.
This gluon-sea interplay is also seen in the other polarized gluon models,
where the smaller (and negative) $\Delta G$ moderate the rise is $g_1^p$.

Extrapolating our results in Fig. 1 to $x=0.002$, we can compare to some of the
models of small-$x$ behavior discussed in Section II. Our moderate gluon
model would give $g_1^p$ a value of about $0.75$ here, which is steeper than
the $A_1$ intercept of $-0.14$ for the isotriplet piece, but not as steep as
the two-gluon model of the Pomeron. It is, however, consistent with the vector
coupling model of Donnachie and Landshoff. The zero polarized gluon model gives
$g_1^p$ a slightly less steep slope, but is also consistent with this model.
Here, the polarized sea dominates $g_1^p$ at small-$x$. The instanton-motivated
gluon model creates a relatively constant behavior for $g_1^p$.

In our treatment, the polarized sea dominates the quark contribution at
small-$x$. Since our basic assumption is $\Delta q/q\sim x$ it follows that
$A_1(x)\sim x$. Therefore, the relation $g_1(x)\approx {{F_2(x)A_1(x)}\over
{2x(1+R)}}$ implies that $F_2$ and $g_1$ have the same behavior at small $x$.
The instanton motivated gluon model gives a constant $g_1^p$ behavior, which
seems to disagree with the apparent rise in $F_2$ and, correspondingly, $g_1$.
Thus, the small-$x$ behavior of the data are not consistent with the negative
$\Delta G$ model. It may therefore be possible to rule out negative
$\Delta G$ at these $Q^2$ values if the error bars on $g_1$ can be reduced
in future PDIS experiments. This does not address the possibility for negative
$\Delta G$ at smaller $Q^2$, where non-perturbative effects are present.

The neutron and deuteron structure functions appear to have more moderate
behavior at small-$x$ (see Figs. 3 and 4). In fact, $g_1^d$ asymptotically
approaches zero, to within experimental errors. Since it is not clear whether
$g_1^n$ is negatively increasing or tending to zero, we cannot conclude
whether there are cancellations of $g_1^p$ and $g_1^n$ at small-$x$ to give
this moderate behavior to $g_1^d$ or whether other nuclear effects could be
present. Similarly, we cannot distinguish between the gluon models with these
data, as readily as the proton case. All of the moderate gluon models seem to
fit the data fairly well. A much larger $\Delta G$ would not provide good
agreement at low-$x$. More precise data at small-$x$ could yield more exact
conclusions.

\mainhead{V. Concluding remarks}

We have constructed a set of flavor-dependent polarized parton distributions
using QCD motivated assumptions and recent PDIS data. The main advantages of
our approach are that: both the Gauge-Invariant (GI) and Chiral-Invariant (CI)
factorization are included, we avoid problems inherent with moments by
performing the entire analysis in $x$-space, the positivity constraint holds
for all flavors and the final form of the parametrizations is easy to implement
for predicting polarized processes. 

There are a number of basic physical assumptions underlying these
distributions. First, the valence distributions are SU(6) motivated and the
$\Delta q_v$ parametrizations are determined using the Bjorken Sum Rule.
The SU(3) sea symmetry is broken due to mass effects in polarizing the heavier
strange quarks. Then, we generate $\Delta q$ from $q$ under well defined 
phenomenological assumptions. Our choice of $\eta(x)$ yields a small-$x$
behavior which is Regge-like and a large $x$ behavior satisfying the counting
rules. We have assumed no unphysical large $\Delta G$ and $L_z$, but have
allowed an explicit interplay between $\Delta S$ and $\Delta G$ via the anomaly
in the CI factorization. The three different polarized gluon models have
different physical bases and provide a reasonable range of possibilities,
which can be narrowed down by future experiments. These gluon models are
consistent with theoretical calculations involving quark models and assumptions
about the orbital angular momentum \cite{bbs,bcd,bj}.

The distributions exhibit success in fitting $g_1^{p,n,d}$ both in $x$
dependence and the integral values: $\int_0^1 g_1^{p,n,d}\>dx$, since these are
built into the parametrizations. Evolution has been performed in LO and NLO,
with little significant difference in the range 1 GeV$^2 \le Q^2\le$ 10
GeV$^2$. Differences start becoming apparent at the $Q^2$ values of other
experiments (around 40-50 GeV$^2$). This will be discussed in more detail in
reference \cite{goram}.

In Section III, we discussed the allowable range of $n$ and $b$ in $\eta(x)$,
subject to the positivity constraint, with $a$ fixed by normalization to data.
These two parameters are tightly constrained together. Thus any variation in
$n$, will restrict the allowable values of $b$, with the most flexibility for
about ${1\over 2}\le n\le 1$. The corresponding possible variation in $b$ is
comparable to the range of $a$ seen in Tables I through III for fixed values
of $b$. The variation in $a$ is primarily due to the different $\eta_{av}$
values, characteristic of the different experimental results. This range is not
significantly large, and since the polarized sea is only a small part of $g_1$,
except perhaps at small-$x$, the differences are not significant to the overall
results we present here.

The results of $g_1^p$ at small-$x$ imply that it may be possible to narrow
down the gluon size with more precise PDIS experiments at small-$x$. Such
experiments are planned at SLAC (E155) and DESY (HERMES). These would also
refine the parametrizations by indicating the behavior of $g_1^i$ at small-$x$.
Comparisons of the $x$-dependent deuteron structure function with the
corresponding proton and neutron structure functions could provide insight
into possible nuclear effects, if they are significant. There are various
possible experiments which would provide a better indication of the size of the
polarized gluon distribution. These include: (1) one and two jet production
in $e-p$ and $p-p$ collisions, \cite{sv,rrs,rs,leg} (2) prompt photon production
\cite{qrrs,gv96,gv94,cm,cg}, (3) charm production \cite{bbg} and (4) pion
production \cite{rs}. Groups at RHIC (STAR), SLAC (E156), CERN (COMPASS) and
DESY (HERA-$\vec {N}$) are planning to perform these experiments in the near
future. For detailed explanations of these experiments, see references 
\cite{gprpw} and \cite{bot}. We are presently calculating the appropriate
processes using the distributions and gluon models presented here \cite{goram}.

{\bf Acknowledgement}: One of us (G.P.R.) would like to thank P. Ratcliffe and
D. Sivers for useful discussions regarding the positivity constraint.

\end{document}